%% file: main.tex
\documentclass[a4paper, amsfonts, amssymb, amsmath, reprint, showkeys, nofootinbib, twoside]{revtex4-1}
\usepackage[english]{babel}
\usepackage[utf8]{inputenc}
\usepackage[colorinlistoftodos, color=green!40, prependcaption]{todonotes}
\input{preamble}
\usepackage[pdftex, pdftitle={Article}, pdfauthor={Author}]{hyperref} 
\begin{document}
\title{Hard gluon evolution in warming medium}

\author{F. G. Ben}
    \email[Correspondence email address: ]{felipe.gregoletto@ufrgs.br}
\author{M. V. T. Machado}
    \affiliation{Universidade Federal do Rio Grande do Sul, Institute of Physics, Porto Alegre, RS, Brazil}
\date{\today} 

\begin{abstract}
We describe the energy distribution of hard gluons travelling through a dense quark-gluon plasma whose temperature increases linearly with time, within a probabilistic perturbative approach. The results were applied to the thermalization problem in heavy ion collisions. In the weak coupling picture this thermalization occurs from ``the bottom up'': high energy partons, formed early in the collision, radiate low energy gluons which then proceed to equilibrate among themselves, forming a thermal bath that brings the high energy sector to equilibrium. We see that, in this scenario, the dynamic we describe must set in around $t \sim 0.5$ fm/c after the collision in order to reach a fully thermalized state at $t \sim 1$ fm/c. We then look at the entropy density and average temperature of the soft thermal bath, as the system approaches (local) thermal equilibrium.
\end{abstract}

\keywords{Quark Gluon Plasma, heavy ion collisions, bottom-up thermalization}

\maketitle

\input{sections/section01.tex}  
\input{sections/section02.tex}
\input{sections/section03.tex}

\input{sections/acknowledgements.tex}

\bibliography{fgbenbib}


\end{document}

%% file: preamble.tex
\usepackage{amsthm}
\usepackage{mathtools}
\usepackage{physics}
\usepackage{xcolor}
\usepackage{graphicx}
\usepackage[left=23mm,right=13mm,top=35mm,columnsep=15pt]{geometry} 
\usepackage{adjustbox}
\usepackage{placeins}
\usepackage[T1]{fontenc}
\usepackage{lipsum}
\usepackage{csquotes}
\usepackage{dcolumn}
\usepackage{bm}

\newcommand{\be}{\begin{equation}}       
\newcommand{\ee}{\end{equation}}       
\newcommand{\beqn}{\begin{eqnarray}}
\newcommand{\as}{\alpha_s}

\newcommand{\ab}{\bar{\alpha}}
\newcommand{\hq}{\hat{q}} 
\newcommand{\kp}{k_{\perp}^2}

%% file: sections/section01.tex
\section{Outline} \label{sec:outline}





Thermalization is of utmost importance in the physics of heavy-ion collisions. The quark-gluon plasma (QGP) formed in the collision is, initially, highly anisotropic and out of equilibrium. As it expands, the QGP undergoes several stages that are characterized by different degrees of freedom and described by different effective theories. In particular, once it has reached a local thermal equilibrium, it is described by relativistic hydrodynamics. In order to reproduce experimental data, however, simulations in relativistic hydrodynamics have to be initiated at early times after the collision, of order $t_{hydro} \sim 1$ fm/c \cite{kurkelaa2016initial}. Understanding how the postcollision debris is able to redistribute energy and reach local thermal equilibrium so quickly has been one of the central topics in the heavy-ion community for the past few years.

In the weak-coupling picture, prethermal evolution undergoes three different stages that lead to thermalization ``from the bottom up'', as presented in the seminal paper in Ref. \cite{baier2001bottom}: first, the high-energy hard gluons from the collision irradiate soft gluons through bremsstrahlung. Those soft gluons carry only a small fraction of the parent gluon's energy, but rapidly increase in number and equilibrate among themselves in the second stage. Third, the system is now formed by a small number of hard gluons, which carry most of the system's energy and travel through a thermal bath of soft gluons. The interaction of the hard gluons with the soft medium is responsible for depositing the energy of the hard sector into the thermal bath, bringing the system to a (local) thermal state. This last stage also sets the time scale $t_{hidro}$, since the first two stages take parametrically less time \cite{kurkela2011bjorken}.

In this work we use a probabilistic approach to describe the evolution of the hard gluons during the last bottom-up stage, taking into account how the average soft-medium temperature $T$ changes during thermalization. 
Our starting point is that the physics of the third bottom-up stage is similar to that of ``jets'' of high momentum travelling through a thermal medium \cite{schlichting2019first}. A series of papers \cite{blaizot2013medium, blaizot2014probabilistic}, based on perturbative QCD, established the formalism for propagation of a high-energy parton in a dense quark-gluon plasma, within the BDMPS-Z framework \cite{baier1997radiative, baier1997radiative2, zakharov1997radiative}. In the original formulation the plasma was considered homogeneous and static. Subsequent work discussed how the formulation can be applied to an expanding medium \cite{adhya2020medium, caucal2020jet}, by using a modified emission rate or by mapping the expanding medium to an effective static one. We applied the available results to the thermalization problem in a previous letter, Ref. \cite{ben2020hard}. 

The main shortcoming to our goal is that previous results, which focus on jet quenching, assume that the high-energy partons travelling through the medium do not alter $T$ in a significant way. Is that assumption valid in the thermalization scheme? During the last bottom-up stage, the temperature of the soft thermal bath increases linearly with time, even during the system's expansion, due to the hard gluons which serve as an energy source \cite{baier2001bottom}. 
In order to address this issue, in this letter we apply the formalism developed in Ref. \cite{blaizot2013medium,blaizot2014probabilistic,caucal2020jet} to hard gluons travelling in a homogeneous quark-gluon plasma whose temperature $T$ increases (linearly) with time. We compare our results to static and Bjorken-expanding mediums, and also use experimental and phenomenological constraints, along with simple thermodynamics, to look at medium properties as the proper time $t$ after the collision approaches $t_{hydro} \approx 1$ fm/c.

%% file: sections/section02.tex
\section{Developments} \label{sec:develop}

Consider the problem of a small number of high energy gluons, of initial energy $E$, travelling through a dense QCD plasma. For now, let us assume that the plasma is homogeneous, static and has a uniform temperature $T$. 
For a parton of high momentum travelling through a QCD medium, the main mechanism of energy loss is in-medium bremsstrahlung, subject to the Landau-Pomeranchuk-Migdal (LPM) suppression, which leads to the BDMPS-Z distribution \cite{baier1997radiative,zakharov1997radiative}. For a thermal medium this was analysed in Ref. \cite{jeon2005energy}, which confirmed that the final distribution is governed by the small number of high energy partons travelling through the medium. 

We shall focus on the (hard) gluon spectrum $D(x,t) \equiv x \dfrac{dN_g}{dx}$, made of $N_g$ gluons of initial energy $E$. The energy $\omega$ of a given gluon is represented by the fraction of the initial energy $x = \omega/E$. In this scheme, a hard gluon will lose energy by democratic branching: first, it emits a particle of small energy $\omega_{br} \ll E$ that lies in the LPM-suppressed region, such that it will split into two gluons of comparable momenta.
Those daughter gluons will cascade further, depositing their energy into the thermal bath of $x \sim T/E$.
Considering $t_{br}$ the typical time scale between successive emissions of energy $\omega \sim \omega_{br}$, the proper time $t$ will be re-scaled as the dimensionless variable
\be
\tau = \bar{\alpha}\sqrt{\frac{\hat{q}}{E}} t = \frac{t}{t_{br}(E)},
\label{eq:taubar}
\ee
where $\ab = \as N_c /\pi$ and $\hat{q}$ is the transverse momentum broadening rate $\hat{q} \equiv dk_{\perp}^2/dt$, or jet quenching parameter. For our purposes, this parameter $\hq$ fully characterizes the interaction of the high energy partons with the medium\footnote{Although we will assume that all relevant medium interactions are encoded in the evolution equation through $\hq$, in general jet-medium interactions are encoded in $\sigma_3(b_{\perp})$ rather than $\hq$. See, for instance, Ref. \cite{zakharov1997radiative,arnold2008high,caron2009g,moore2020transverse}.}. 
In other words, we will assume that $\hq$ contains the information of the medium properties and its evolution. For a static medium, $\hq = \hq_0$ is a constant, but it must be modelled accordingly in different scenarios.

Considering that different branchings are independent of each other, Refs. \cite{blaizot2013medium,blaizot2014probabilistic} derived an evolution equation for the gluon distribution,
\be
\frac{\partial D(x,\tau)}{\partial \tau} = \int \mathcal{K}(z,\tau) \bigg[\sqrt{\frac{z}{x}} D(\frac{x}{z}, \tau) - \frac{z}{\sqrt{x}}D(x,\tau)\bigg] \mathrm d z.
\label{eq:evol}
\ee
The initial condition is normalized to a single gluon of energy $E$, i.e., $D(x,0) = \delta(1-x)$. Also, the kernel function $\mathcal{K} (z, \tau)$ is related to the emission spectrum $I(z,\tau)$ of the parton through
\be
\bar{\alpha} \mathcal{K}(z,\tau) = \frac{\mathrm{d}I}{\mathrm{d}z \mathrm{d}\tau},
\ee
and it depends on the medium and its evolution \cite{arnold2009simple, adhya2020medium}. The factor of $\ab$ was extracted in order to make our notation consistent with the one used in Ref. \cite{arnold2009simple}.

When the medium is homogeneous and static, $\mathcal{K} (z, \tau)$ is in fact independent of $\tau$,
\begin{eqnarray}
\label{eq:K} 
\mathcal{K}(z) &=& \frac{f(z)}{[z(1-z)]^{3/2}} = \mathcal{K}(1-z), \\
f(z) &=& [1 - z(1-z)]^{5/2},
\end{eqnarray}
valid in the limit $z \ll \tau^2$. Notice that the connection of Eq. \eqref{eq:evol} to the medium properties is only through Eq. \eqref{eq:taubar}, i.e., through the mapping of $\tau$ to the ``physical'' time $t$.
We also recall that Ref. \cite{blaizot2013medium} showed that, if you make the additional simplification $f(z) = 1$, Eq. \eqref{eq:evol} can be solved analytically, leading to
\be 
D_0(x,\tau) = \frac{\tau}{\sqrt{x}(1 - x)^{3/2}} \mathrm{e}^{\frac{-\pi\tau^2}{1-x}}.
\label{eq:Danal}
\ee

We now move beyond the static plasma. We shall follow an approach that is a faithful description only for relatively soft medium-induced emissions. Such emissions have very short formation times, which implies that they can be treated as instantaneous and independent of each other, as required by Eq. \eqref{eq:evol}. Ref. \cite{caucal2020jet} points out that, by the same argument, the emission rate for a non-static medium is taken to be the same as that for a static medium, replacing $\hq$ by an instantaneous $\hq(t)$ at the emission time. Following these assumptions, the scaled time $\tau$ in Eq. \eqref{eq:taubar} will be replaced\footnote{See Ref. \cite{caucal2020jet} for a complete derivation. For a simple hand-waving motivation, think of the leading parton going through a non-static medium as of going through successive layers of static plasmas, each of different properties represented by a different $\hq_i$, in time intervals $\Delta t_i$. By Eq. \eqref{eq:taubar}, after going through many layers $\Delta \tau = \sum_i \ab \sqrt{\dfrac{\hq_i}{E}} \Delta t_i$, from which Eq. \eqref{eq:tau} follows.} by
\be
\tau \equiv \int_{t_o}^t \ab \sqrt{\frac{\hq(t)}{E}} \, \mathrm{d} t.
\label{eq:tau}
\ee

The behavior of $\hq(t)$ with time encodes the medium evolution. During its emission, a gluon picks up transverse momentum squared $k_{\perp}^2 \sim n_s \as^2 \Delta t_F$, where $\Delta t_F$ is the gluon formation time and $n_s$ the density of soft gluons \cite{baier2001bottom}. From this, we are able to estimate $\hat{q} \sim k_{\perp}^2 / \Delta t_F \sim \as^2 n_s$ (see, for instance, Ref. \cite{iancu2018jet} for a comprehensive discussion on $\hq$ in different scenarios). Using that $n_s \sim T^3$, in general $\hq$ goes like $\hq \sim n_s \sim T^3$. For a static medium, it follows that $\hq = \hq_o$ is a constant. Considering that the medium undergoes uniform (isentropic) longitudinal expansion, as initially proposed by Bjorken \cite{bjorken1983highly}, $n_s \sim 1/t$ (since the volume increases linearly with time). This leads to $\hq (t) = \hq_o (\dfrac{t_o}{t})$ for Bjorken expansion. The parameters $t_o$ and $\hq_o$ will be discussed shortly.

In this letter, however, we are interested in a different scenario: a medium whose temperature $T$ is uniform, but increases linearly with time, i.e., $T \sim t$. We shall also assume that $T$ rises slowly enough that the soft thermal bath that comprises the medium can be considered to be in thermal equilibrium at all times. We keep an eye in our goal to describe the third bottom-up stage: Ref. \cite{baier2001bottom} estimates (parametrically) that $T$ rises linearly, due to the hard gluons that act as energy sources. 

Since the linear rise in $T$ is such a central assumption in our analysis, let us briefly take a step back and see where does it come from in this framework. During its emission in a dense medium, a gluon picks up transverse momentum squared $\kp$ due to interactions with the medium. This introduces a lower bound  $\kp \gtrsim \hq \Delta t_F$, where $\Delta t_F \approx 2\omega/\kp$ is the gluon formation time, which implies $\Delta t_f \lesssim \sqrt{2\omega/\hq}\equiv t_c(\omega)$.
Recall that Eq. \eqref{eq:evol} follows from treating multiple emissions as a probabilistic branching process, in which the branching rate is governed by the BDMPS-Z spectrum \cite{blaizot2014probabilistic}. The probability distribution $dP_{\mathrm{BDMPSZ}}$ is well approximated by the formula \cite{caucal2020jet}
\be
\omega \frac{dP_{\mathrm{BDMPSZ}}}{d\omega} \simeq \ab \frac{L}{t_c(\omega)} = \ab \sqrt{\frac{\omega_c}{\omega}},
\label{eq:prob}
\ee
where $\omega_c = \hq L^2/2$, and $L$ is the distance traveled by the parton inside the medium.
The typical energy loss happens for $\omega \frac{dP}{d\omega} \gtrsim 1$, which requires $\omega \lesssim \ab^2 \omega_c \equiv \omega_{br}.$ From Eq. \eqref{eq:prob}, $\omega \lesssim \omega_{br}$ implies $L \gtrsim t_c(\omega)/\ab \equiv t_{br}(\omega)$. In summary, the typical emission has $\omega \simeq \omega_{br}$ and takes time of order $t_{br}$, i.e., $t_{br}$ is the typical time to make an emission of $\omega \simeq \omega_{br}$.

We now want to estimate how the flow of energy from the hard sector affects the temperature in this framework. The emission rate is of order $dN_{br}/dt \sim n_h/ t \sim Q_s^2/\as t^2$, where $n_h \sim Q_s^2/\as  t$ is the density of hard gluons \cite{baier2001bottom}. Taking $ t \sim t_{br}$, the energy-flow rate is given by
\be
\omega_{br} \frac{dN_{br}}{dt} = \omega_{br} \frac{Q_s^2}{\as t_{br}^2}.
\ee
From $\kp \sim n_s\as^2\Delta t_F$, along with $\Delta t_F \sim \omega/\kp$, we also have
\be
\Delta t_F^2 \sim \frac{\omega}{n_s \as^2}.
\ee
Since $\Delta t_F \sim t_c(\omega)$, we use $t_{br} = t_c/\ab$ to obtain $t_{br} \sim \dfrac{\omega^{1/2}}{n_s^{1/2}\as^2}$. From here on, we see that this estimate converges with the one presented in Sec. II of Ref. \cite{baier2001bottom}. The energy-flow rate becomes
\be
\omega_{br} \frac{dN_{br}}{dt} \sim \as^3 Q_s^2 n_s \sim \as^3 Q_s^2 T^3,
\ee
as $n_s \sim T^3$. From $d\epsilon/dt \sim d(T^4)/dt \sim \as^3Q_s^2 T^3$, it follows that $T \sim t.$ We see that the probabilistic treatment of the emissions, as governed by Eq. \eqref{eq:prob}, which also leads to Eq. \eqref{eq:evol}, naturally leads to the linear behavior of $T$. We point out that this behavior is not a direct consequence of Eq. \eqref{eq:evol}, but rather a more general result that comes directly from the underlying branching rate. The connection of Eq. \eqref{eq:evol} and Eq. \eqref{eq:prob} can be made clearer if we notice that, as presented in Ref. \cite{blaizot2013medium}, Eq. \eqref{eq:prob} may be written as $\dfrac{dP}{dzd\tau} = \dfrac{1}{2}\dfrac{\mathcal{K}(z)}{\sqrt{x}}$, where $\mathcal{K}(z)$ goes into the description of the hard evolution through Eq. \eqref{eq:evol}. 

For the warming medium, the relation $\hq \sim T^3 \sim t^3$ suggests
\be
\hq(t) = \frac{\hq_o}{t_o^3} t^3,
\label{eq:qt}
\ee
where $t_o$ stands for the initial time and $\hq_0 = \hq(t_o)$. For our purposes, $t_o$ represents the time when the third bottom-up stage is initiated. Taking $t = 0$ at the collision, $t_o$ is some time in the interval $0 \lesssim t_0 < t_{hydro}$. Phenomenologically, $t_o$ may be treated as a free parameter to be varied in simulations, subject to $t_{hydro} \sim 1$ fm/c. Also, $\hq_o$ may be treated as another free parameter in the theory or it may be set by imposing $\hq_o t_o \approx Q_s^2$, as in Ref. \cite{caucal2020jet}. In order to keep the number of degrees of freedom to a minimum, we shall take $\hq_o = Q_s^2/t_o$ in what follows.

The scaling relation Eq. \eqref{eq:tau} will lead to a different expression for $\tau(t)$ in each case. For a static medium,
\be
\tau_{ST} = \ab \sqrt{\frac{\hq_o}{E}}(t - t_o).
\ee
while for Bjorken expansion we have
\be
\tau_{BJ} = \ab \sqrt{\frac{\hq_o t_o}{E}} \ln{(\frac{t}{t_o})},
\ee
and for $\hq(t)$ given by Eq. \eqref{eq:qt} we have
\be
\tau_T = \ab \frac{2}{5 \, t_o^{3/2}} \sqrt{\frac{\hq_o}{E}} (t^{5/2} - t_o^{5/2}).
\ee

A key feature of this approach, based upon the one in Ref. \cite{caucal2020jet}, is that the assumption of soft emissions $\omega \ll \omega_c \equiv \hq(L)L^2/2$ (where $L$ represents the medium length scale), with the scaling variable $\tau$ given by Eq. \eqref{eq:tau}, implies that $\mathcal{K}(z)$ is still given by Eq. \eqref{eq:K}, independent of $\tau$. Scaling violations due to harder emissions will not be considered in this letter, but they are expected to be small. For future work, $\mathcal{K}(z,\tau)$ could be treated more precisely (see, for instance, Ref. \cite{adhya2020medium, ben2020hard, arnold2009simple}).
In this case, the solution to Eq. \eqref{eq:evol} in terms of the scaling variable, $D(x,\tau)$, is independent of $\hq(t)$. Solutions differ in terms of the physical time $t$ through the mapping from $\tau$ to $t$. To be more precise,  $D(x,\tau)$ is obtained from Eq. \eqref{eq:evol}, and then $D(x,t) = D(x, \tau(t))$, with $\tau(t)$ given by Eq. \eqref{eq:tau} in each case\footnote{A more complete treatment could use $\mathcal{K}(z,\tau)$ as defined in Ref. \cite{arnold2009simple}, with the $\tau$ given by Eq. \eqref{eq:taubar}. In that case, $\mathcal{K}(z,\tau)$ depends on $\hq(t)$ through a differential equation.}.

\begin{figure*}[!htb]
\centering
\includegraphics[scale=0.7]{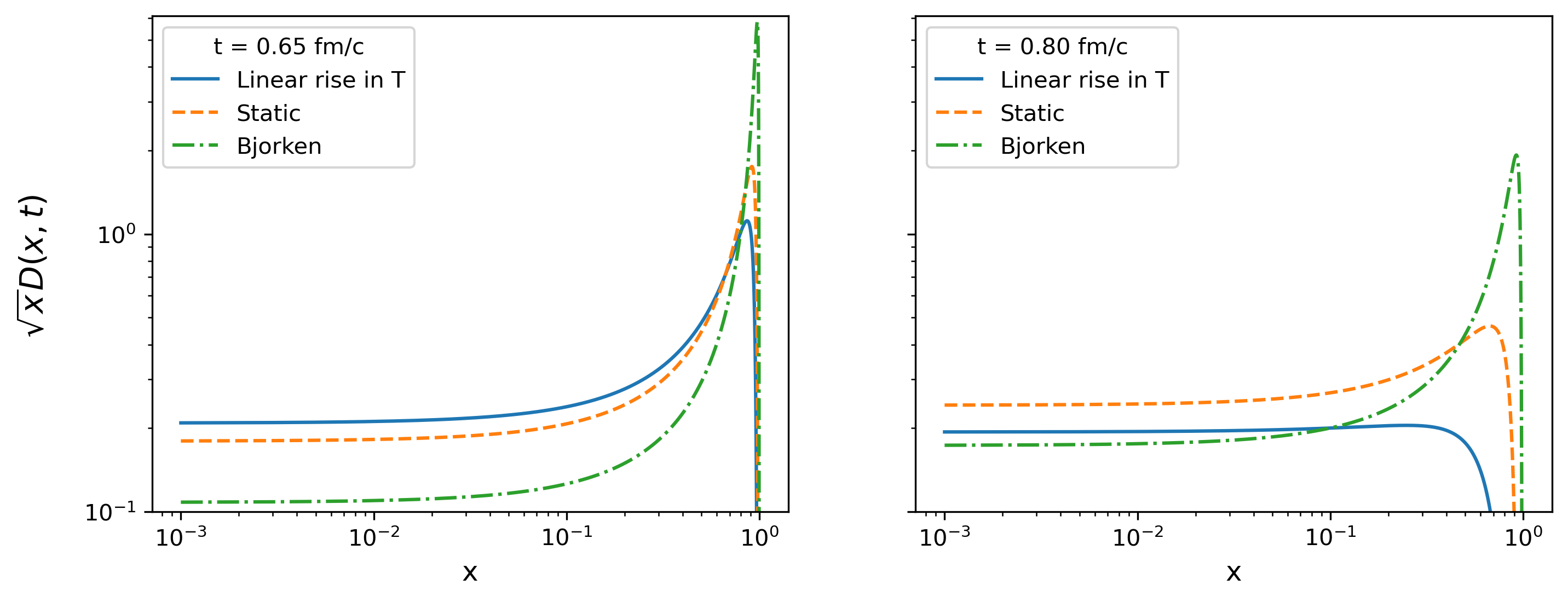}
\caption{Graph of $\sqrt{x}D_o(x,t)$ for the three cases: static, Bjorken (isentropic) expansion and linear rise in $T$, at different values of $t$.}
\label{fig:spectrum}
\end{figure*}

Let us begin our discussion by comparing different scenarios in the light of the additional simplification $f(z) = 1$, which leads to the analytic solution Eq. \eqref{eq:Danal}. We shall take $E \sim Q_s = 2$ GeV (since hard gluons still carry an energy of order $Q_s$ at the beginning of the third bottom-up stage \cite{baier2001bottom}) and $t_o = 2.5$ GeV$^{-1} \approx 0.5$ fm/c (for now, the reader may think of this as a numerical value to make the comparison, but the physical motivation behind this choice will be explained shortly). The spectrum $\sqrt{x} D(x,t)$ is plotted as a function of $x$ in Fig. \ref{fig:spectrum} for three different cases: static, Bjorken (isentropic) expansion and linear rise in $T$ (which accounts for expansion together with how the hard gluons affect the temperature as they travel through the soft bath). First, notice the general features of the distribution $D(x,t)$: the behavior alludes to a source, initially located at $x = 1$, that is dampened as the spectrum propagates in the direction of $x \rightarrow 0$. Comparing the evolution at different times, notice that, although free expansion makes the evolution slower than the static case, expansion \textit{plus warming} makes the evolution faster. We should not be surprised, since $\hq \sim T^3$. The rise in $T$ suggests that the effect of the hard energy sources on the medium is able to compensate for the dilution from the expansion, bringing the system close (but faster) to the behavior of the static case. Clearly, this is an interesting result on its own, but specially so in the thermalization problem: the effect of warming makes energy fade out from the hard sector faster.

\begin{figure}[!htb]
\centering
\includegraphics[scale=0.6]{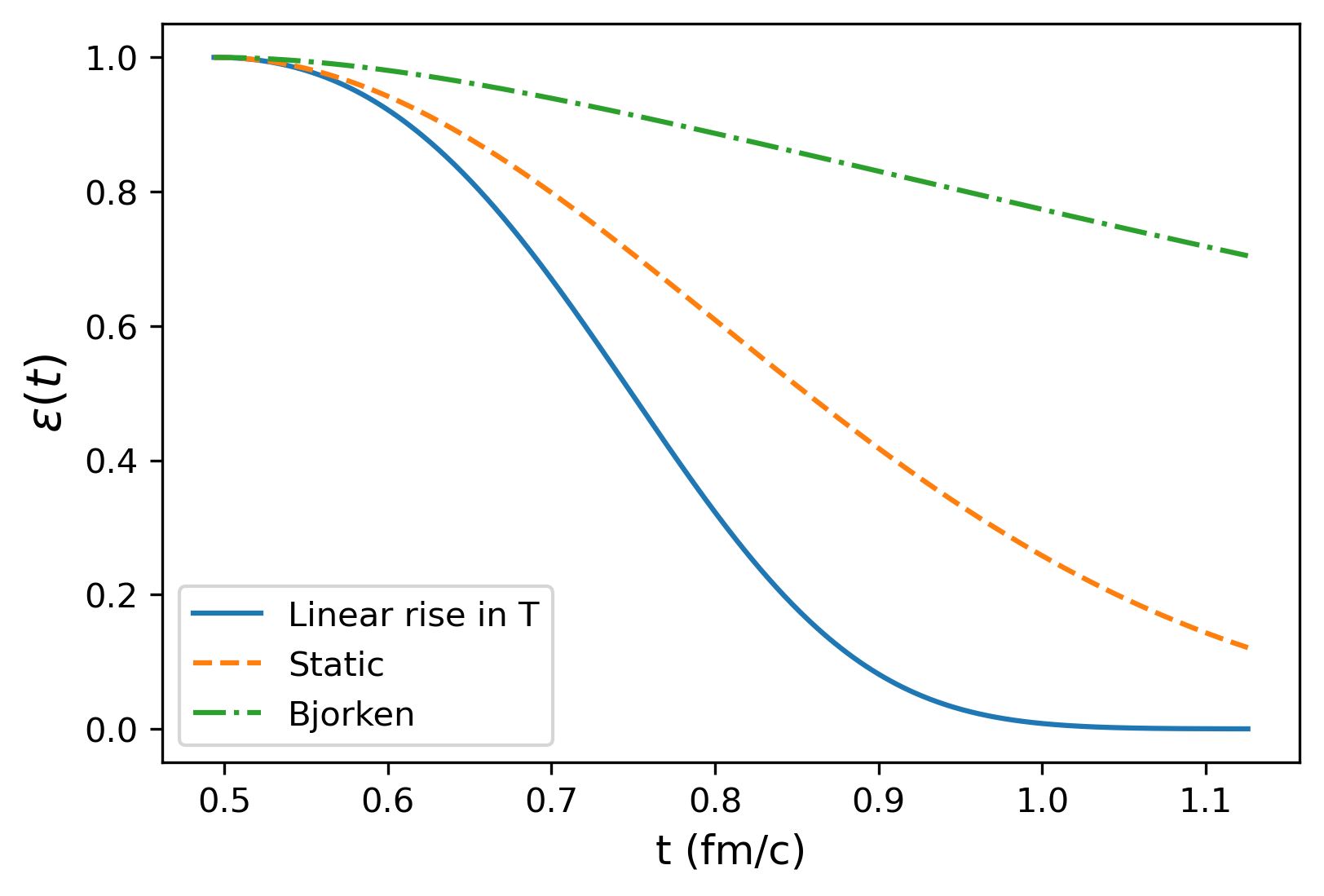}
\caption{$\mathcal{E}(t)$, the fraction of the initial energy contained in the hard sector, for the three cases -- static, Bjorken (isentropic) expansion and linear rise in $T$ -- as a function of time.}
\label{fig:energyfraction}
\end{figure}

The fraction of the initial energy contained in the hard spectrum is given by
\be
\mathcal{E}(\tau) \equiv \int_0^1 D(x,\tau) \, \mathrm{d}x = \int_0^1 x \frac{dN_g}{dx}  \, \mathrm{d}x .
\label{eq:energy}
\ee
Another general feature of the dynamics from Eq. \eqref{eq:evol} is that $\mathcal{E}(\tau)$ decreases with time. Formally, this apparent violation of energy conservation is due to the singularity at $x = 0$, which acts as a ``drain''
to the energy of higher modes. Physically, energy from the hard sector is being deposited in the thermal bath of $x \sim T/E$. The entire system (hard gluon and plasma) may be treated as thermal once $\mathcal{E}$ goes to zero, i.e., we expect $\mathcal{E}(t = t_{hydro}) \approx 0$. For the analytic solution $D_o(x,\tau)$ in Eq. \eqref{eq:Danal}, we have that
\be
\mathcal{E}_0(\tau) = \int_0^1 D_0(x,\tau) \, \mathrm{d}x = \mathrm{e}^{-\pi\tau^2},
\label{eq:Eanal}
\ee
which implies an exponential decrease in energy. Fig. \ref{fig:energyfraction} shows a comparison for $\mathcal{E}(t)$ as a function of the physical time $t$ for the same cases presented in Fig. \ref{fig:spectrum}. The value of $t_o$ was set to $t_o = 2.5$ GeV$^{-1} \approx 0.5$ fm/c in order to obtain $\mathcal{E}(t_{hydro} = 1 \mathrm{fm/c}) \approx 0$ in the third case, although it could be kept as a free parameter to be varied in the simulations and treated as a systematic uncertainty. Physically, it means that, in this simplified model, our dynamic must set in around $t_o = 0.5$ fm/c after the collision in order to reach a fully thermalized state at $t_{hydro} = 1$ fm/c.

Instead of considering only numerical values, it is also interesting to consider the relationship of $t_o$ with $t_{hydro}$ from the analytic solution in Eq. \eqref{eq:Eanal}. Taking  $\mathcal{E}(t_{hydro}) = c$, and collecting leading terms, it follows that $t_o \sim A^{-1/4}\ab^{1/2} Q_s^{1/4} t_{hydro}^{5/4}$, where $A \equiv [-\log(c)\frac{25}{4\pi}]$. The numerical value of the parameter $A$ depends on the particular value of $c$ that one is willing to consider small enough, such that $\mathcal{E}(t_{hydro}) = c \approx 0$. 
Parametrically, Ref. \cite{baier2001bottom} estimated that $t_{hydro} \sim \as^{-13/5} Q_s^{-1}$, along with the estimate that the third bottom up stage sets up at a time scale $t_o \gg \as^{-5/2} Q_s^{-1}$.
Using the same parametric estimate for $t_{hydro}$, our approach leads to $t_o \sim A^{-1/4}\as^{-11/4} Q_s^{-1}$, which is greater than $\as^{-5/2} Q_s^{-1}$ by a factor of order $(A\as)^{-1/4}$.
Taking $c = 0.01$ in order to make a simple numerical estimate (which corresponds to assuming that $\mathcal{E}(t)$ fell bellow $1\%$), we would find $t_o/t_{hydro} \sim 0.7$.
We recall that these are parametric estimates from leading terms, while $t_o = 2.5$ GeV$^{-1}$ follows from the (complete) analytic solution.

\begin{figure}[!htb]
\centering
\includegraphics[scale=0.6]{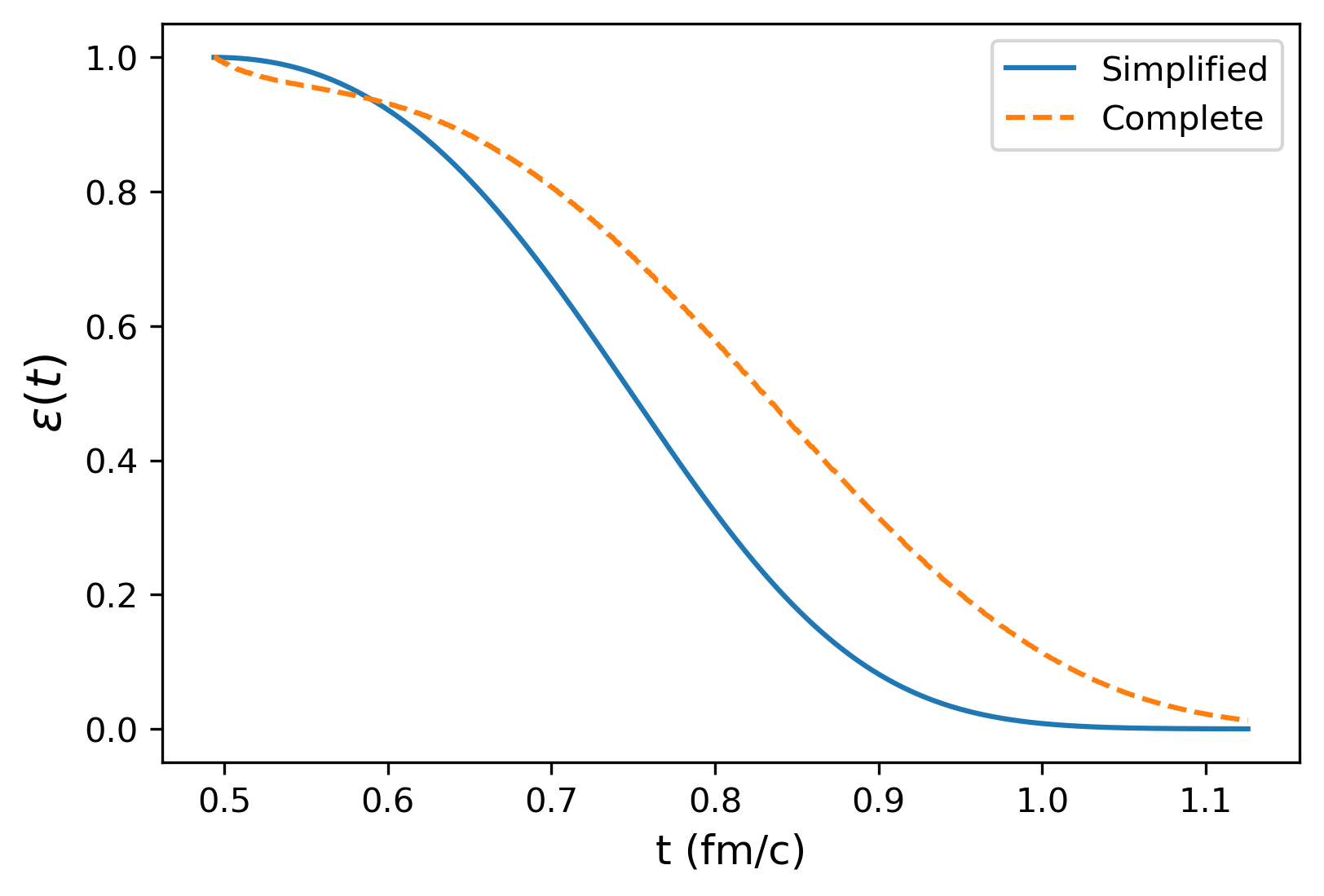}
\caption{$\mathcal{E}(t)$ as a function of time with the additional simplification $f(z) = 1$ (solid line) and with the complete $\mathcal{K}$ in Eq. \eqref{eq:K} (dashed line), for the case of linear rise in $T$.}
\label{fig:energyfraction2}
\end{figure}

We have also solved Eq. \eqref{eq:evol} numerically, using the complete $\mathcal{K}(z)$ in Eq. \eqref{eq:K}. A comparison to the analytic solution (which pertains to the case of $f(z)=1$) in Eq. \eqref{eq:Eanal} is presented in Fig. \ref{fig:energyfraction2} for the case of linear rise in $T$. All parameter values were kept the same as in previous plots for comparison. Using the complete kernel, energy fades out from the hard sector a little slower. We have verified that this would change our estimate $t_o$ to $t_o = 2$ GeV$^{-1}$ in order to reach $\mathcal{E}(t_{hydro} = 1 \mathrm{fm/c})\approx 0$. The assumption of $f(z)=1$ overestimates the emission rate of the spectrum for intermediate values of $z$, such that the complete kernel leads to slower thermalization, which in turn requires the final-stage dynamic to set in earlier in order to reach thermalization in a reasonable time scale.

In summary, the energy loss of a high-energy parton in a weakly-coupled plasma may be described analytically using a probabilistic approach, as seen in recent results in the context of jet quenching physics. The thermalization problem in heavy-ion collisions, as described in the seminal paper in Ref. \cite{baier2001bottom} through kinetic theory, goes through a similar scenario, as high-energy partons travel through a thermal bath and thermalization occurs as those hard partons deposit its energy in the medium, leading to a linear rise in temperature during this process. 
In this Letter, we use the probabilistic approach to look at the evolution of the hard sector, through an evolution equation that follows directly from treating multiple emissions as a probabilistic branching process. We take into account the linear rise in $T$ in the medium description, and we find that this dynamic, corresponding to the third bottom-up stage, must set in at around $t_o \sim 2 - 2.5$ GeV$^{-1} \approx 0.5$ fm/c in order that a thermalized system is reached by $t_{hydro} = 1$ fm/c.

%% file: sections/section03.tex
\section{Medium properties} \label{sec:conclusions}

In this section, we shall describe how general features of the quark-gluon plasma evolve in this simplified model. 
In the picture we have established, hard gluons travel through a quark-gluon plasma of uniform temperature $T$, whose value increases linearly with time as the hard sector deposits its energy into the plasma. In particular, we are interested in how the entropy density $s$ of the QGP evolves as the system approaches $t_{hydro}$.
Most experimental observables in a heavy-ion collision are not very sensitive to pre-thermal evolution, but entropy production is a notable exception \cite{schlichting2019first}: once the system is described by \textit{ideal} hydrodynamics, the expansion becomes nearly isentropic, which means that entropy production in a heavy-ion collision must be governed by pre-thermal evolution. In other words, a measure of the final state entropy is able to provide us with an estimate for the entropy density at $t_{hydro}$. Ref. \cite{hanus2019entropy} determined the entropy per unit rapidity produced in Pb-Pb collisions at the LHC, and estimated that, for the 0\% - 10\% most central Pb-Pb collisions at $\sqrt{s_{NN}} = 2.76$ TeV, 
\be
s_h = s(t_{hydro}) = 82.3 \, \mathrm{fm}^{-3}.
\ee

For an ideal gas of massless, non-interacting constituents, the entropy density is given by
\be
s(T) = \frac{2\pi^2}{45}\nu(T) T^3,
\label{eq:sdens}
\ee
where $\nu$ counts the number of bosonic degrees of freedom plus $7/8$ times the number of fermionic degrees of freedom \cite{muller2005entropy}. An ideal gas of non-interacting gluons and three flavors of massless quarks has $\nu = 47.5$. However, QCD thermodynamics does not describe the quark-gluon plasma as an ideal gas of non-interacting particles, except at infinite temperatures. As presented in Ref. \cite{muller2005entropy}, lattice QCD shows that, for finite temperatures, $\nu(T)$ varies over a wide range of values. Considering the crossover temperature $T_c = 170 \pm 10$ MeV, in the range $2T_c < T < 5T_c$ we have that $\nu(T)$ is between $70\%$ and $80\%$ of that for an ideal quark-gluon plasma, that is, $ 33 < \nu < 38$ \cite{karsch2002lattice}, whereas for $T \approx 200$ MeV we have $\nu \approx 25$. For what follows, we shall take $25 < \nu < 38$ and treat $\nu$ as a theoretical uncertainty. From $s_h = 82.3$  fm$^{-3}$ and Eq. \eqref{eq:sdens}, we have
\be
T_h = T(t_{hydro}) \approx 358 \, \mathrm{MeV},
\ee
close to the value of $T_h \approx 340$ MeV obtained in Ref. \cite{hanus2019entropy}.

\begin{figure}[!htb]
\centering
\includegraphics[scale=0.6]{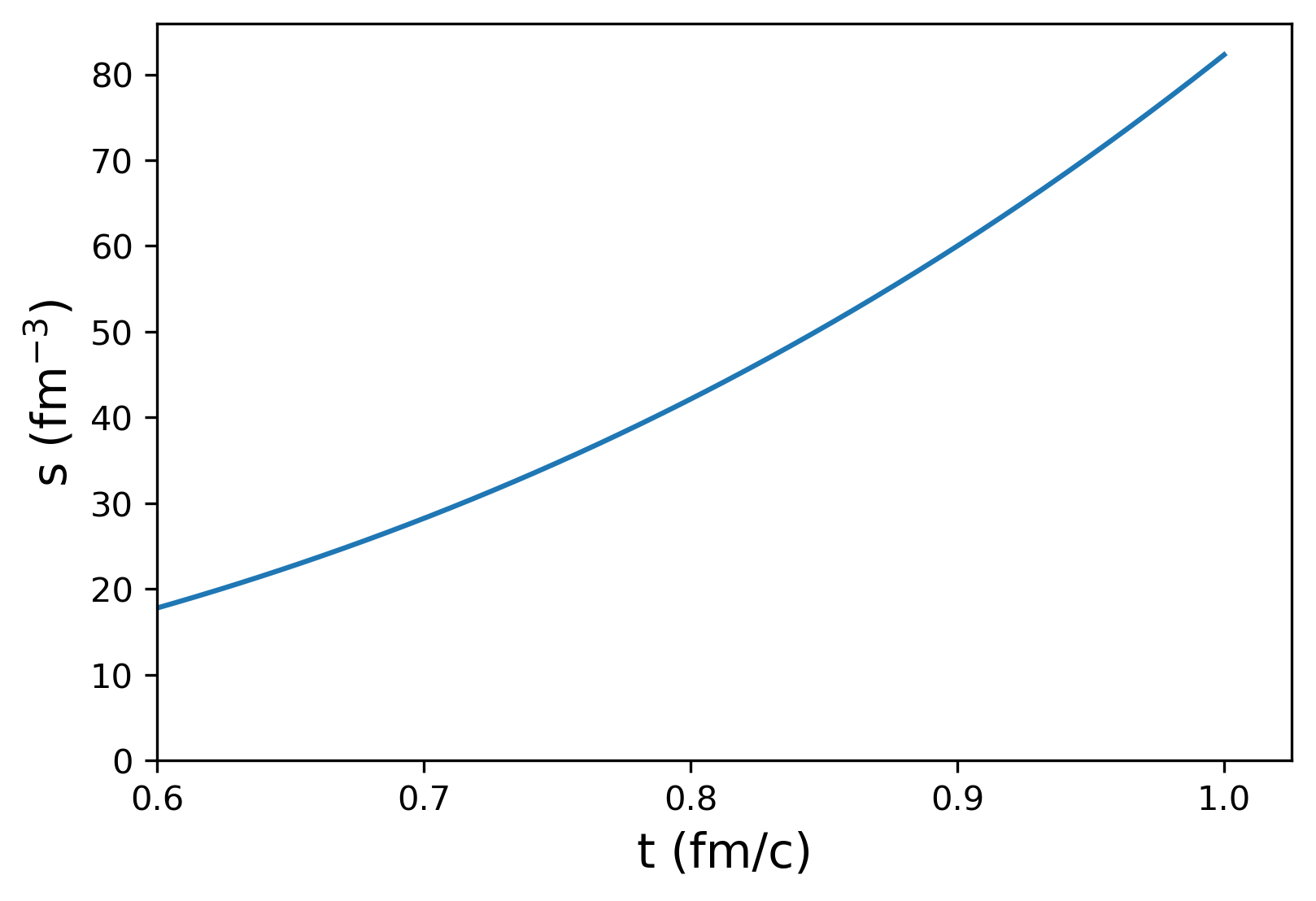}
\caption{Entropy density of the thermal bath as a function of time.}
\label{fig:entropy}
\end{figure}

\begin{figure}[!htb]
\centering
\includegraphics[scale=0.58]{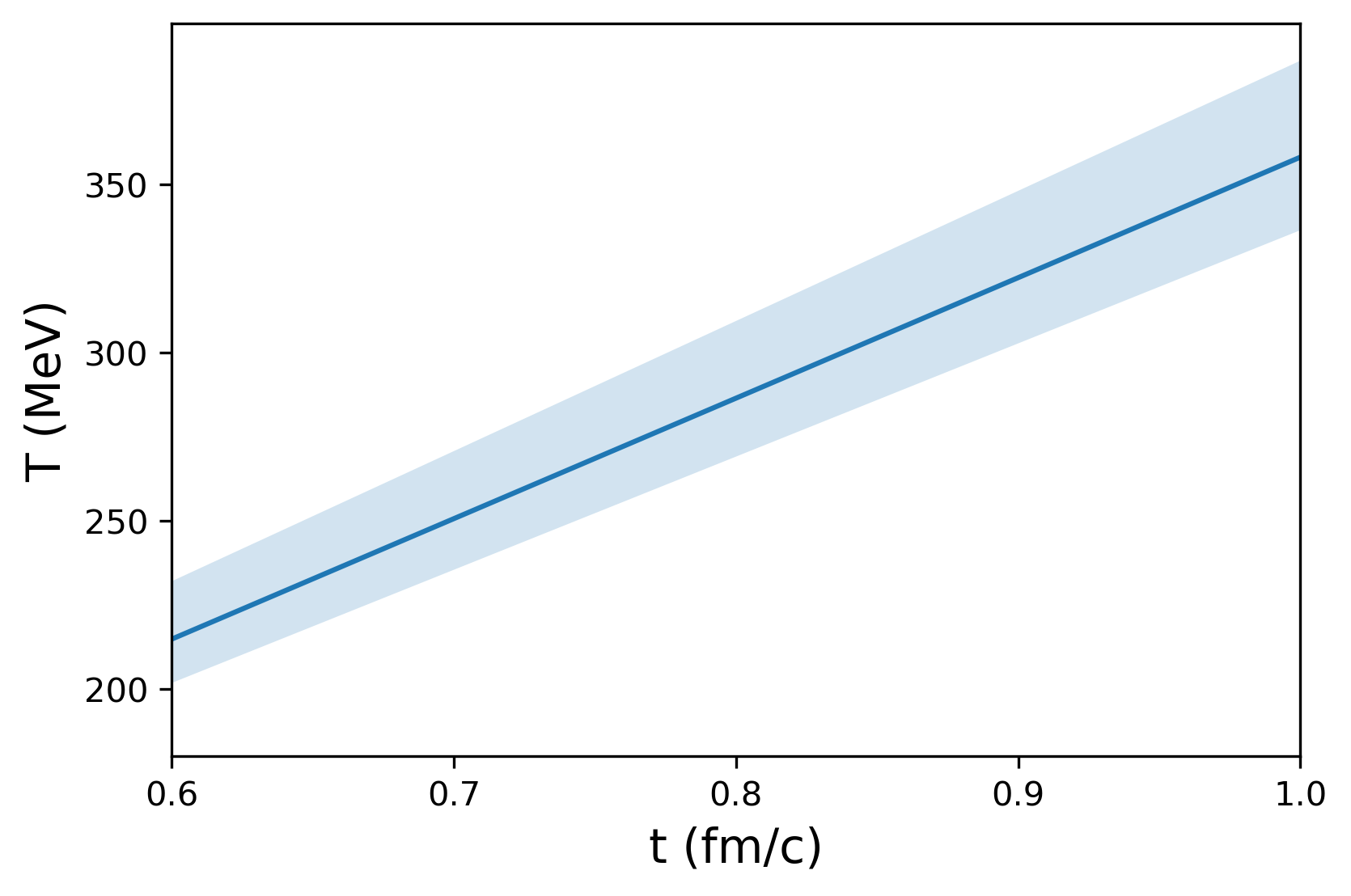}
\caption{Average temperature as a function of time. The uncertainty band represents the theoretical uncertainty in the number of degrees of freedom $25 < \nu(T) < 38$.}
\label{fig:temperature}
\end{figure}

Let us now attempt to describe how those values are approached in the interval $t_o < t \lesssim t_{hydro}$. From Eq. \eqref{eq:sdens}, with $T \sim t$, it follows that $s(t) \sim t^3$ for the thermal bath, where the value $s(t_{hydro})$ is fixed by $s_h$. For $t_o < t < t_{hydro}$ this is a lower bound on the entropy density of the entire system, since we are not taking into the account the contribution from the hard sector. Going back to Eq. \eqref{eq:sdens}, we are also able to estimate $T(t)$ from $s(t)$. Fig. \ref{fig:entropy} and \ref{fig:temperature} present the entropy density and the temperature of the medium as a function of time. Notice that the time evolution follows directly from $s \sim t^3$ and does not depend directly on $t_o$. The variation of $T$ with $\nu(T)$ is plotted as an uncertainty band in the value of $T$. Notice, also, that the initial entropy density is less than a third of the final one. This initial entropy comes essentially from the decoherence of the initial gluon field, while the remaining entropy is generated during thermalization \cite{fries2009decoherence}. Ref. \cite{iida2014time} studied entropy density from decoherence as a function of time in heavy-ion collisions (see, also, Ref. \cite{Tsukiji:2016krj, Tsukiji:2017pjx}) and obtained a value of $15 < s_{dec} <20$ fm$^{-3}$ for $0 < t < 2$ fm/c, which is in agreement with our lower bound: from Eq. \eqref{eq:sdens}, $s(t_o) \approx 10$ fm$^{-3}$ for $t_o = 0.5$ fm/c (from the simplified kernel), while $s(t_o) \approx 5$ fm$^{-3}$ for $t_o = 0.4$ fm/c (when using the complete branching kernel). 

%% file: sections/acknowledgements.tex
\section*{Acknowledgements} \label{sec:acknowledgements}
The authors thank Niels Schlusser for discussion. This work was financed by the Brazilian funding agency CNPq and in part by CAPES - Finance Code 001.